\documentclass[twocolumn,prl,aps,superscriptaddress,showpacs]{revtex4-1}

\usepackage{gensymb}
\usepackage{graphicx}
\usepackage{xcolor}
\usepackage{bm,hyperref}

\begin{document}

\title{Collapse of ferromagnetism and Fermi surface instability near reentrant superconductivity of URhGe}

\author{A. Gourgout}
\affiliation{Univ. Grenoble Alpes, INAC-PHELIQS, F-38000 Grenoble, France}
\affiliation{CEA, INAC-PHELIQS, F-38000 Grenoble, France}
\author{A. Pourret}
\email[E-mail me at: ]{alexandre.pourret@cea.fr}
\affiliation{Univ. Grenoble Alpes, INAC-PHELIQS, F-38000 Grenoble, France}
\affiliation{CEA, INAC-PHELIQS, F-38000 Grenoble, France}
\author{G. Knebel}
\affiliation{Univ. Grenoble Alpes, INAC-PHELIQS, F-38000 Grenoble, France}
\affiliation{CEA, INAC-PHELIQS, F-38000 Grenoble, France}
\author{D. Aoki }
\affiliation{Univ. Grenoble Alpes, INAC-PHELIQS, F-38000 Grenoble, France}
\affiliation{CEA, INAC-PHELIQS, F-38000 Grenoble, France}
\affiliation{Institute for Materials Research, Tohoku University, Oarai, Ibaraki, 311-1313, Japan}
\author{G. Seyfarth}
\affiliation{Univ. Grenoble Alpes, LNCMI, F-38042 Grenoble Cedex 9, France}
\affiliation{CNRS, Laboratoire National des Champs Magn\'etiques Intenses LNCMI (UJF, UPS, INSA), UPR 3228, F-38042 Grenoble Cedex 9, France}\author{J. Flouquet}
\affiliation{Univ. Grenoble Alpes, INAC-PHELIQS, F-38000 Grenoble, France}
\affiliation{CEA, INAC-PHELIQS, F-38000 Grenoble, France}

\date{\today }

\begin{abstract}

 We present thermoelectric power and resistivity measurements in the ferromagnetic superconductor URhGe for magnetic field applied along the hard magnetization {\it b}-axis of the orthorhombic crystal. Reentrant superconductivity is observed near the the spin reorientation transition at $H_{R}$=12.75~T, where a first order transition from the ferromagnetic to the polarized paramagnetic state occurs. Special focus is given to the longitudinal configuration, where both electric and heat current are parallel to the applied field. The validity of the Fermi-liquid $T^2$ dependence of the resistivity through $H_R$ demonstrates clearly that no quantum critical point occurs at $H_R$. Thus the ferromagnetic transition line at $H_R$ becomes first order implying the existence of a tricritical point at finite temperature. The enhancement of magnetic fluctuations in the vicinity of the tricritical point stimulates the reentrance of  superconductivity. The abrupt sign change observed in the thermoelectric power with the thermal gradient applied along the {\it b}-axis together with the strong anomalies in the other directions is a definitive macroscopic evidence that in addition a significant change of the Fermi surface appears through $H_R$.
\end{abstract}

\pacs{71.18.+y, 71.27.+a, 72.15.Jf, 74.70.Tx}
%\keywords{}

\maketitle

%The last years have seen growing interest in quantum phase transitions (QPTs), which have a strong impact in explaining several exotic low-temperature properties of a variety of materials, which do not fit conventional many-body theories.Moreover, even though a QPT is strictly speaking a zero-temperature instability, its experimental manifestations are observed at finite temperature within a rather wide temperature region as a function of a non-thermal control parameter.

Quantum phase transitions (QPT) are a central topic in contemporary condensed matter research. Their rich underlying physics plays an important role in explaining exotic low-temperature properties of a variety of strongly correlated materials like high-T$_{C}$ superconductors \cite{He2014}, quantum magnets \cite{Sachdev2008} or heavy-fermion compounds \cite{Lohneysen2007, Si2010}. Strictly speaking, a QPT is a zero-temperature instability, yet its manifestations can be observed at finite temperature within a rather wide temperature region as a function of a non-thermal control parameter. Recent theoretical \cite{Belitz2005, Mineev2011, Schmalian2004,Yamaji2006,Imada2010,Chubukov2004} analyses of ferromagnetic (FM) QPTs have shown that generally the second order phase transition turns into a first-order one at a tricritical point (TCP) in the proximity of a FM QPT when approaching the absolute zero temperature in clean systems. Experiments in FM systems such as ZrZn$_2$ \cite{Uhlarz2004} or UGe$_2$ \cite{Taufour2010} confirm this trend. However, in other systems (such as YbNi$_4$P$_2$ \cite{Steppke2013}) a continuous second order QPT has been invoked. In principle, a control parameter can be tuned opportunely in order to move the TCP to zero temperature, generating a quantum critical end point (QCEP). Some compounds are located close to a QCEP at ambient conditions \cite{Grigera2001, Plakhty2003,Giovannetti2011,Taufour2010}. 

In the present paper we study the magnetic phase diagram of the orthorhombic Ising-type ferromagnet URhGe and its interplay with superconductivity (SC) \cite{Aoki2001}. URhGe is one of the four uranium based compounds, besides UGe$_2$ \cite{Saxena2000}, UCoGe \cite{Huy2007a}, UIr \cite{Akazawa2004}, where microscopic coexistence of ferromagnetism and  SC has been observed. In URhGe, the magnetic moments $M_0\approx0.4\mu_B$ are oriented along its easy {\it c}-axis. A transverse magnetic field higher than the superconducting critical field $H_{c2}$ applied along the hard magnetization {\it b}-axis induces at low temperature a reorientation of the magnetic moments from {\it c} to {\it b} axis \cite{Hardy2011} at  $H_R=11.75~$T. A field reentrant superconducting phase (RSC) appears in a narrow field window around $H_R$ below $T_{RSC}=410~$mK \cite{Levy2005}. It has been suggested that the transverse magnetic field tunes the system in the vicinity of the  TCP \cite{Levy2007}. Thus it is a key case to study a FM QPT. It allows to investigate the interplay of magnetic fluctuations and possible Fermi surface (FS)  changes with SC.

Thermoelectric power (TEP) is an excellent probe to detect electronic singularities and FS changes notably in strongly correlated electron systems as it is sensitive to the derivative of the density of states and the electronic scattering with respect to the energy at the Fermi energy \cite{Miyake2005b}. Pertinent exemples are heavy-fermion compounds such as CeRu$_2$Si$_2$ \cite{Boukahil2014,Pfau2011}, CeRh$_2$Si$_2$ \cite{PalacioMorales2015}, YbRh$_2$Si$_2$ \cite{Pfau2013,Pourret2013b}, or URu$_2$Si$_2$ \cite{Pourret2013a}. Here, we present systematic TEP and resistivity measurements on URhGe with different orientations of thermal current $J_Q$ and electric current $J_e$ with respect to the magnetic field, which is always applied along the {\it b}-axis. Experimental details are given in the Supplemental Material. We will focus on the longitudinal response with currents and field along the {\it b}-axis. Special attention is given on the temperature dependence of the resistivity at various magnetic field. The validity of the Fermi-liquid $T^2$ dependence through $H_R$ demonstrates clearly that no QCP occurs at $H_R$, thus the FM transition line at $H_R$ becomes first order implying the existence of a TCP at finite temperature. Evidence of a first order transition at $H_R$ was reported by torque \cite{Levy2009}, Hall resistivity \cite{Aoki2014}, and has been recently confirmed by NMR experiments \cite{Kotegawa2015}. The abrupt variation in the TEP for the three directions of $J_Q$  at $H_R$ is a macroscopic signature of a drastic change of the FS. Previous signatures had been detected by quantum oscillations \cite{Yelland2011} and Hall effect \cite{Aoki2014} experiments.

\begin{figure}[h!]
	 \begin{center}
		\includegraphics[width=0.9\linewidth]{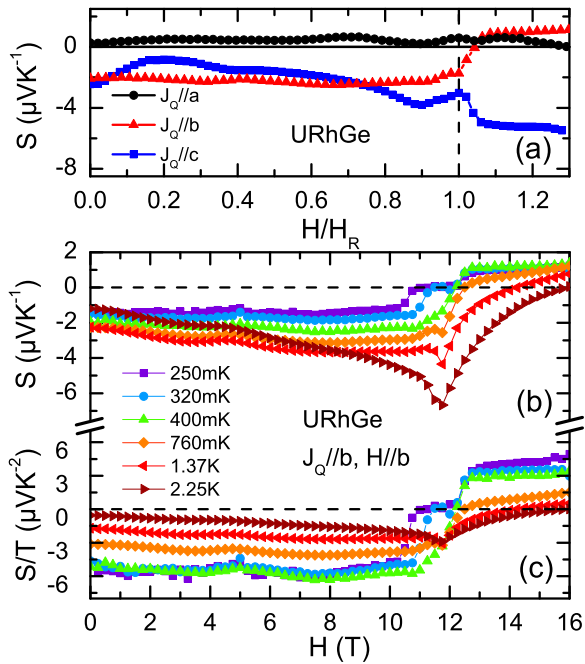}
		\caption{\label{Fig1} (Color online). (a) Thermoelectric power $S$ as a function of magnetic field $H$ along the {\it b}-axis normalized by $H_R$ at $T\approx$ 470~mK for $J_Q$ along the three crystallographic directions. (b) the TEP $S$ and (c) $S/T$ at different temperatures from 250~mK to 2.25~K for $J_Q \parallel b$ and $H\parallel b$.}
	\end{center}
\end{figure}

Figure~\ref{Fig1}(a) shows the field dependence along the {\it b}-axis of the TEP for $J_Q$ along the three main crystallographic directions at $T=470$~mK, just above the critical temperature of the RSC state ($T_{RSC}=410~$mK). The TEP is clearly anisotropic and shows very pronounced anomalies at $H_R$ for the {\it b} and {\it c} direction. For $J \parallel a$, although the TEP is always positive and small, it shows small anomalies around $H_R$. In this direction the signature in the TEP of the scattering term is suspected to be small as $J_Q$ stays perpendicular to the direction of the magnetic moments even above the reorientation at $H_R$. For the transverse configuration, $J_Q \parallel c$,  the TEP is always negative and decreases with increasing field. It shows a clear peak at $H_R$. In the longitudinal configuration, $J_Q \parallel b$, the TEP is always negative in the FM state, has a step-like transition at $H_R$ and becomes positive in the polarized paramagnetic (PPM) state above $H_R$. As already reported in Hall resistivity experiments \cite{Aoki2014}, small anomalies occur around 1.5~T and 5~T suggesting minor changes in the FS. Without any orbital effect in the longitudinal configuration, the TEP change at $H_R$ originates most likely from a FS reconstruction as suggested previously \cite{Yelland2011, Aoki2014}. We will now focus on the results $J_Q \parallel b$ with $H\parallel b$.

 \begin{figure}[h!]
	 \begin{center}
		\includegraphics[width=0.9\linewidth]{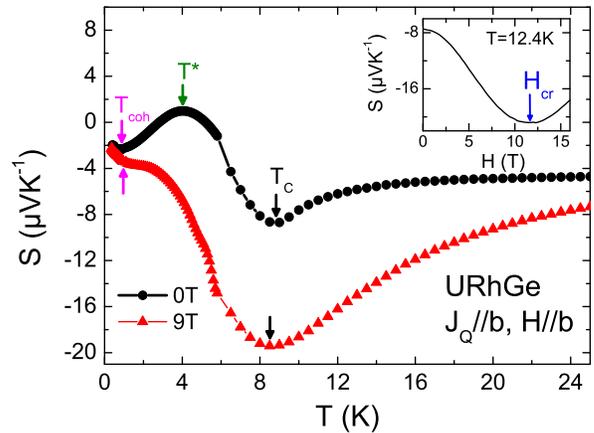}
		\caption{\label{Fig2} (Color online). Temperature dependence of the TEP between 1~K and 25~K for $H=0$ and 9~T. The black arrows mark $T_C$, the green and pink arrows indicate the position of anomalies labeled $T^*$ and $T_{coh}$ respectively. In the inset, the field dependence of $S$ at 12.4~K (above $T_C$) shows a broad minimum indicated by the blue arrow at $H_{cr}$.}
	\end{center}
\end{figure} 

The magnetic field dependence of the TEP, $S(H)$, and the TEP normalized by temperature, $S/T(H)$, from 250~mK to 2.25~K for $J_Q$ and $H \parallel$ {\it b}-axis is represented in Fig.~\ref{Fig1} (b) and (c), respectively. $S$ is negative below and positive above $H_R$. At 2.25~K, $S$ shows a sharp negative peak at $H_R = 11.75~T$. With decreasing temperature the transition becomes sharper and finally step-like. At 250~mK $S$ shows a two step transition with $S$=0 from 10.5~T to 12.5~T indicating the presence of the RSC in this system around $H_R$. In a simple two-band picture, the sign of the TEP is set by the product of the effective mass and the mean free path of the heat carriers \cite{Miyake2005b}. Therefore, the observation of $S/T$ with different signs below and above $H_R$ (see Fig. \ref{Fig1}(c)), implies that the nature of this pocket changes across the transition. While we cannot identify individually the pockets participating in this transition, the result finds a natural explanation if one assumes that the suppression of the FM state is accompanied by a substantial reconstruction of the FS without changing the compensated nature of the system.  We can also notice (see Fig. \ref{Fig1}(c)) that at $H_R$ for $T >T_{RSC}$, $S/T$ is temperature independent  with a value of -2.8$ \mu$VK$^{-2}$, indicating that the electronic singularity in the density of states occurs at a peculiar value of the entropy per carrier. 

\begin{figure}[h!]
	 \begin{center}
		\includegraphics[width=0.9\linewidth]{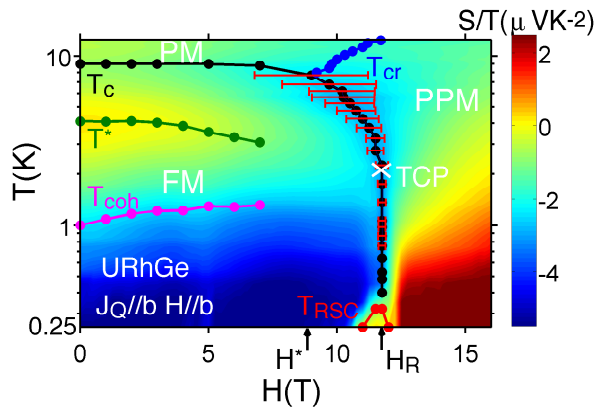}
		\caption{\label{Fig3} (Color online). Linear color map of $S/T$ in the ($T$,$H$) plane. The Curie temperature $T_C$ (black circles), the energy scales $T^*$ (green circles) and $T_{coh}$ (pink circles),  the reentrant superconductivity $T_{RSC}$ (red circles) and the crossover line $T_{cr}$ between the PM and the PPM state (blue circles) are superimposed. The transition width observed in the TEP around $H_R$ is also represented (red horizontal lines).}
	\end{center}
\end{figure} 

Figure~\ref{Fig2} displays the temperature dependence of the TEP for $H=0$ and 9~T. With decreasing temperature, a first minimum occurs around the Curie temperature $T_C \approx 9.5~K$. Inside the FM state, two other anomalies appear at $T^*\approx4$~K and $T_{coh}\approx 1$~K. $T^*$ may mark a characteristic energy of the interplay between magnetic excitations and the establishment of the FM FS below $T_C$. $T_{coh}$ indicates the entrance in the coherent low temperature Fermi-liquid regime in which the TEP is linear in $T$ for $T \rightarrow 0$~K. In the inset, a typical field dependence of the TEP in the paramagnetic (PM) state at $T  =12.4$~K$>T_C$ is represented. The TEP still exhibits a broad minimum around $H_{cr}\approx$12~T defining a crossover $T_{cr}(H)$ between the PM and PPM state. This crossover can still be observed at 36~K and 18~T.

\begin{figure}[h!]
	 \begin{center}
		\includegraphics[width=0.9\linewidth]{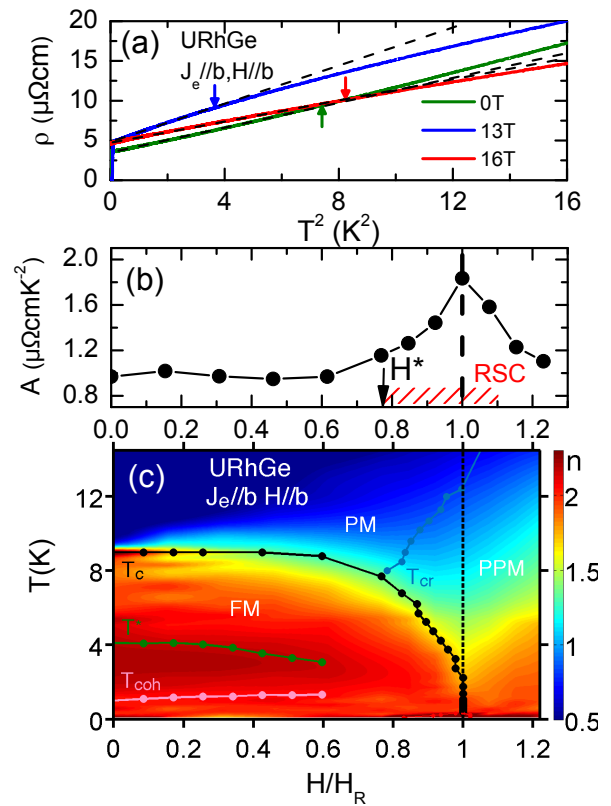}
		\caption{\label{Fig4} (Color online). (a) Resistivity as a function of $T^2$ for $J_e\parallel b$, $H\parallel b$ below 4~K for different magnetic fields. Linear fits at low temperature are represented by dashed lines. The vertical arrows indicate the deviation from $T^2$ dependence. (b) Field dependence of the $A$ coefficient of the resistivity. (c) Linear color map of the exponent $n$ of the resistivity ($\rho(T) = \rho_0 + A T^{n}$)  in the ($T$, $H/H_{R}$) plane. The different anomalies observed in the TEP are superimposed on the phase diagram. }
	\end{center}
\end{figure} 

Figure~\ref{Fig3} presents $S/T$ as a color plot in the ($T$,$H$) plane. We can clearly see that at low temperature $S/T$ is strongly negative (dark blue) in the FM state (below $H_R$) and becomes positive (dark red) in the PPM state. The different anomalies obtained in the TEP measurements for $J_Q$, $H \parallel b$ are superimposed. The width of the FM transition (for details see Fig.~S1 of the Supplemental Material) observed in the $H$ scans of the TEP around $H_R$ is also represented (red horizontal lines).  The sudden increase of the transition width when increasing temperature is a clear signature of crossing the TCP, which hence can be located precisely at $T_{TCP}$=2~K and $H_{TCP}$=11.5~T. Concomitantly,  the low temperature energy scales $T^*$ and $T_{coh}$ seem to converge to the same point in the ($T$,$H$) plane, suggesting a link with the TCP. Magnetic torque measurements located a TCP at 11.45~T \cite{Levy2006, Levy2009} for a perfect alignment along the {\it b}-axis leading exactly to the same value of $H_R$. For $T<T_{TCP}$, the FM transition becomes first order and is independent of field. 

The temperature dependence of the resistivity ($\rho$) is represented as a function of $T^2$ in Fig.~\ref{Fig4}(a). At very low temperature $\rho(T)$ follows the Fermi-liquid theory with $\rho(T) = \rho_0 + AT^2$. With increasing temperature $\rho (T)$ deviates from the $T^2$ dependence with an exponent $n<2$ for all fields except for $H=0$. We fitted $\rho(T)$ such as $\rho(T)=\rho_{0}+AT^n$, on a sliding window of 400 mK below 14~K. $\rho_0$ is the residual resistivity and $A$ the coefficient characterizing the amplitude of the inelastic scattering.  The field dependence of $A$ determined at lowest temperature is shown in Fig. \ref{Fig4}(b). It exhibits a peak at $H_R$, indicating an increase of the effective mass associated to spin fluctuations. Similar behavior of $A(H)$ has been observed in the transverse configuration \cite{Miyake2008}. The enhancement in $A(H)$ starts roughly near the characteristic field where the crossover line $T_{cr}(H)$ intercepts $T_{C}(H)$ at $H^*=8.8$~T (black arrow in Fig. \ref{Fig4}(b)) and where $T_C(H)$ starts to decrease. Astonishingly the magnetization along the {\it c}-axis, $M_c$, starts to decrease already at $H^*$ \cite{Levy2005}, see Fig. S2 of Supplemental Material. RSC in the TEP and in the magnetoresistance measurements is found at 270~mK between 10 and 12.5~T. The strong enhancement of $A$ in the field range 8-15~T with a maximum at $H_R$ is in excellent agreement with the observation of the enhancement of nuclear relaxation rates $\frac{1}{T_1}$ and $\frac{1}{T_2}$ detected by NMR \cite{Kotegawa2015, Tokunaga2015}. We notice that our $T_{TCP}$ estimation is lower than that proposed in Ref. \onlinecite{Kotegawa2015} where $T_{TCP}\approx$~4~K.

\begin{figure}[h!]
	 \begin{center}
		 \includegraphics[width=0.9\linewidth]{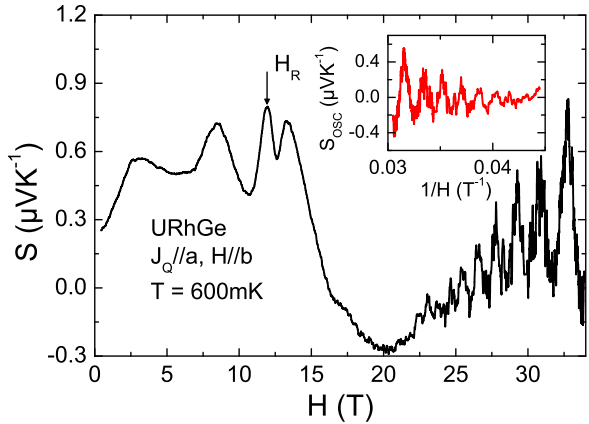}
		\caption{\label{Fig5} Magnetic field dependence of  TEP in URhGe for $J_Q \parallel a$, $H\parallel b$  up to 34~T at 600~mK. $S$ shows small anomalies around $H_R$ and quantum oscillations above 22~T, represented in $1/H$ in the inset.}
	\end{center}
\end{figure} 

A linear color plot of the exponent $n$ of the resistivity in the ($T$,$H/H_R$) plane is represented in Fig. \ref{Fig4}(c) where the different anomalies observed in the TEP are superimposed. Remarkably, below 2~K n$\approx$2 is found to be field independent and thus, no quantum critical behavior appears. This is in excellent agreement with the first order transition below the TCP close to $H_R$. We find $n \approx$ 2.3 around 4~K where the anomaly $T^*$ has been observed by TEP. This observation of $n>$2 inside the FM state could be related to magnetic excitations.

The data reported in the different phase diagrams lead to an unambiguous determination of the position of the TCP of  the FM to PPM transition, which is characterized by the {\it c} to {\it b} axis switch of the magnetization. Signatures of  FS instabilities at the FM to PPM transition are clearly observed in the field variation of $S(H)$ through $H_R$. Furthermore, Hall effect \cite{Aoki2014} as well as ARPES experiments \cite{Fujimori2014} point out a FS change on crossing the PM-FM phase at $T_C$ in low field on cooling. Thus three different FS will correspond to the PM, FM and PPM phases. The possibility of a Lifshitz transition at $H_R$ in URhGe was proposed in  Ref.~\onlinecite{Yelland2011} from Shubnikov de Haas (SdH) measurements performed at an angle of 10$\degree$ from the {\it b}-axis to escape from the RSC \cite{Yelland2011}. For this angle, no first order transition and no RSC, but a crossover at $H_R$ is expected. The experimentally observed SdH oscillations below $H_R$, corresponding to a small orbit of only a few percent of the Brillouin zone, vanish on approaching $H_R$. A possible explanation is the collapse of the orbit. It is claimed that this Lifshitz-type transition, leading to the collapse of the Fermi velocity, is the driving force for RSC. However, as shown in Fig.~\ref{Fig5}, the TEP in URhGe for $J_Q \parallel a$, $H\parallel b$ shows large quantum oscillations above 22~T, represented as a function of $1/H$ in the inset. The corresponding frequency, $\approx 500$~T, is very similar to the frequency observed in the SdH measurements. Hence a Lifshitz transition as the sole driving force for RSC seems unlikely. In our study the misalignment is always less than 1$\degree$ and the transition just above the RSC is clearly first order and thus it cannot be of sole Lifshitz nature. Instead, we give macroscopic evidence that RSC is associated with both, a FS instability and critical fluctuations when $T_{C}(H)$ vanishes. Surprisingly, neglecting the FS change, excellent agreement is found in the description of RSC as a function of magnetic field and pressure in a crude phenomenological model where the enhancement of $A(H)$ reflects the enhancement of the effective mass and hence of the superconducting coupling constant \cite{Miyake2008,Miyake2009}. An open question remains the field dependence of the FS inside the dome of RSC and whether this dome can be described with a unique FS. 

%Combining previous results with our new TEP and resistivity measurements

% FS instabilities occur on entering in the new ordered phase either FM or PPM. Pressure and angular studies lead to propose that the QCEP is respectively near 1.5~GPa\cite{Miyake2009} and for $P=0$ at an angle of 5$\degree$ from {\it b} axis corresponding to {\it c} magnetic field compound of 1~T. Just in the vicinity of QCEP, RSC collapses.
 
%The analysis of the $A$ coefficient of the $AT^2$ Fermi liquid term of the resistivity shows. 

Recently RSC was described in a Landau approach taking into account a two bands approach due to the splitting of the bands in the FM domain \cite{Mineev2015} in the FM domain and stressing the importance of longitudinal fluctuations. In agreement with our experiments, it is shown that in a transverse field ($H\parallel$ {\it b}) the PM-FM transition switches from second to first order at a TCP ($H_{TCP}$, $T_{TCP}$) close to $H_R$. The optimum of $T_{RSC}(H)$ is predicted to be roughly half of $T_{TCP}$. In our experiment $T_{TCP}\approx 2$~K and $T_{RSC}\approx$0.4~K, hence $T_{RSC}\approx T_{TCP}/5$. Furthermore, as observed experimentally, $T_{RSC}(H)$ is expected to fall down asymmetrically on both sides of $H_R$. The predicted decrease of $M_c \propto \sqrt{T_C(H)}$ cannot be properly tested due to the lack of accuracy of the existing magnetization data. The renormalized spin-fluctuation theory \cite{Morya} predicts $M_c(H)$ varying as $T_C^{\frac{2}{3}}$ for the collapse of the FM state at a QCP. The vicinity of $H_{TCP}$ from $H_R$ ($H_{TCP}/H_R\approx$ 0.97) is close to what is observed in UGe$_2$ under pressure ($P_{TCP}/P_C\approx$ 0.96).

To summarize, we present clear evidences that on top of a large enhancement of the fluctuations detected here and very recently in NMR experiments \cite{Kotegawa2015, Tokunaga2015}, FS instabilities occur at $H_R$. These fluctuations associated to energy scales converging to the TCP very close to $H_R$ confirm the first order nature of the transition and the absence of a QCP. The role of longitudinal and transversal fluctuations observed close to $H_R$ on the RSC is still under debate. Another interesting proposal is that soft magnons could possibly generate a new attractive pairing interaction for the RSC \cite{Hattori2013}. It is worthwhile to notice that the interplay of FS instabilities and SC is a quite challenging question as a similar problem remains unsolved for high-T$_{C}$ materials as well as for the 115-Ce compounds \cite{Knebel2011}. The additional ingredient of FS instabilities in strongly correlated electronic systems, where the interaction of the quasiparticles themselves is responsible of the superconducting pairing, deserves clearly theoretical treatment. A further experimental challenge is to clarify possible differences in superconducting phases on both sides of  $H_R$.

%Recent NMR studies \cite{Kotegawa2015} indicate that the magnetic fluctuations along the b axis and the c axis develop at around the first-order spin reorientation. In frame of the Landau phenomenological theory it was found that phase transition between anisotropic FM and PM states under strong enough magnetic field perpendicular to direction of easy magnetization changes from the second to the first order type defining a TCP. The critical magnetic fluctuations near the TCP have been put forward as a source stimulating  reentrant superconductivity \cite{Mineev2015}. TCP can never reach the zero temperature or simply be dislocated far enough from the superconducting region on the phase diagram. Reentrant superconductivity in URhGe can arise even in absence of critical fluctuations due to drastic increase of longitudinal susceptibility in vicinity on the TCP \cite{Mineev2015}. The strong change of sign in the TEP and the independent value of $S/T$ at $H_R$ indicate clearly a reconstruction of the Fermi surface in the vicinity of the TCP which has never been taken into account in theoretical calculations.

\begin{acknowledgments}
%\textbf{5. Conclusion}

The authors thank F. L\'evy, K. Behnia and V. Mineev for stimulating discussions. This work has been supported by the French ANR (project PRINCESS), the ERC (starting grant NewHeavyFermion), ICC-IMR, KAKENHI, REIMEI, the EuromagNET II (EU contract no. 228043), LNCMI-CNRS is member of the European Magnetic Field Laboratory (EMFL).
\end{acknowledgments}

\end{document}